\documentclass{sig-alternate-05-2015}
\usepackage{booktabs}
\usepackage{enumitem}
\usepackage{balance}
\usepackage{xcolor}
\usepackage{listings}
\makeatletter
\def\@copyrightspace{\relax}
\makeatother

\begin{document}

\title{KeyGuard: Using Selective Encryption to Mitigate Keylogging in Third-Party IME}

\numberofauthors{2}  
\author{

\alignauthor
Jia Wang\\
       \affaddr{Computing and Software Systems}\\
       \affaddr{University of Washington Bothell}\\
       \affaddr{Bothell, WA, USA}\\
       \email{jiawang@uw.edu}
\alignauthor
Brent Lagesse\\
        \affaddr{Computing and Software Systems}\\
        \affaddr{University of Washington Bothell}\\
        \affaddr{Bothell, WA, USA}\\
        \email{lagesse@uw.edu}
}

\maketitle
\begin{abstract}
As mobile devices become ubiquitous, people around the world have enjoyed the convenience they have brought to our lives. At the same time, the increasing security threats that rise from using mobile devices not only have caught attention from cyber security agencies but also have become a valid concern for mobile users. Keylogging is one of the mobile security threats caused by using insecure third-party IME (input method editor) applications. Keylogging, as the name suggests, keeps track of user\rq s key events performed on the device and stores all the events in a log. The log could include highly sensitive data such as credit card number, social security number, and passwords. This paper presents a novel solution by intercepting the keystroke events triggered by a user and encrypting them before sending them to the third-party IME, making the third-party IME unable to log what the users actually entered on the screen. Input will be decrypted when showing on text view on the underlying app. This solution addresses the fundamental reason why an IME may leak sensitive information since an IME will no longer have access to the user\rq s actual sensitive information, which will greatly reduce the chance of leaking sensitive information by using a third-party IME while maintaining the functionalities of the third-party IME at the same time. 
\end{abstract}



\section{Introduction}
At the beginning of year 2014, the use of Internet on mobile devices has exceeded desktop for the first time in history in the US \cite{bib:pdg2012}. Computing systems have transitioned from requiring users to have highly in depth technical knowledge to use; however modern advances in computing, particularly with mobile devices, have allowed users to use systems without fully understanding the enabling technologies. Given the ubiquitous use of mobile devices to access Internet, security concerns in those emerging environments have arisen as well. A questionnaire research \cite{mohsen2016investigating} showed concerning result that a noticeable percentage of smartphone users don\rq t fully understand the technical details and/or security status of his/her smartphones, i.e. one\rq s android phone could be already rooted when he/she doesn\rq t believe so and neither does he/she understand what the ramifications could be.

The security concerns in mobile devices are real and potentially affecting every mobile user. Per a study conducted by McAfee \cite{mcafee}, security vulnerabilities have been detected in several famous apps that are widely used by the general public, notably Costco app sending login credential in plain-text, Sogou IME (input method editor) collected and uploaded data of user\rq s personal mobile device to its server and Sina and Weibo (two popular social networking apps in China) sent private chat conversation in plain-text. Given the fact apps developed by major well-established businesses still contain potential security vulnerabilities, the majority of smartphone users may not be aware of potential security vulnerabilities. Thus, preventing security vulnerabilities (such as keylogging) in emerging environments is crucial in offering a safe mobile device user experience. 

The impact of security threats in mobile devices can be vast. Almost every people in the US has at least one mobile device. People use and even rely on mobile devices everyday but most people do not have sufficient security awareness. Security threats in mobile device range from malicious apps, spams and phishing to botnet. One of the security threats in mobile device is known as keylogging, which logs all the keyevents on the mobile device and could potentially leak the users\rq sensitive information. This paper is attempting to identify and solve this particular threat in mobile devices.

The paper explores a novel solution - \textit{KeyGuard} - by intercepting the keystroke events triggered by a user and encrypting them before sending them to the third-party IME, making the third-party IME unable to know what the user actually entered on the screen. It will decrypt the encrypted text so that the original text is shown on text view on the underlying app. It attempts to resolve the root cause of leakage of sensitive information by preventing IME from having access to sensitive information at all. In contrast, some of the existing solutions tried to explore reactive measures in the case sensitive information or partial sensitive information had already been exposed to the IME. The solution proposed will address the fundamental reason why an IME may leak sensitive information since an IME will no longer have access to the user\rq s actual sensitive information, which will greatly reduce the chance sensitive information leakage by using a third-party IME.

\section{Background}
\subsection{Keylogging Threat}
Keylogging, as the name suggests, keeps track of user\rq s key events performed on the device and dumps all the events into a log. The log could include highly sensitive data such as credit card number, social security number, passwords to various resources like online banking and e-commerce websites, email addresses and personal contact numbers. While a benign app may want to remember and log all the key events triggered by a user and upload the texts a user has entered to its remote servers on cloud to improve text suggestion accuracy and provide personalized experience, if those data were transferred as plain text, then any man in the middle can sniff and intercept the sensitive information. Not to mention in the case of a malicious app, it would be collecting sensitive information about a user on purpose to prepare to launch an attack, which may cause significant loss to the user, physically and/or financially.

\begin{figure}[ht]
  \graphicspath{{fig/}}
  \begin{center}
  \includegraphics[scale=0.3]{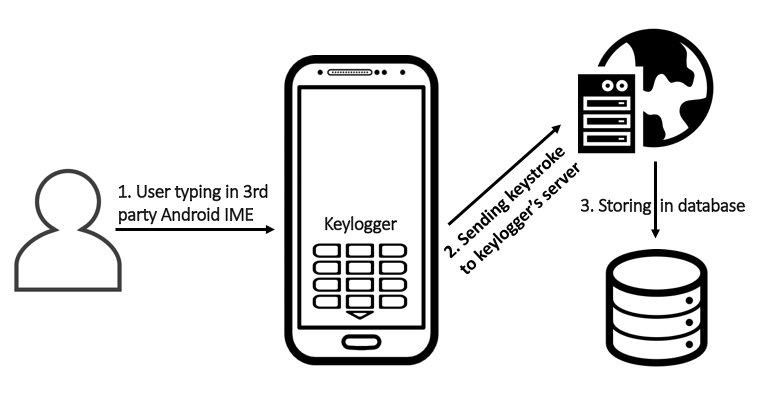} 
  \caption{A typical keylogging scenario.}
  \label{fig:scenario}
  \end{center}
  \end{figure}

A typical scenario of keylogging is shown in Figure \ref{fig:scenario}. When user enters text on his/her mobile device through a malicious third-party IME, the malicious third-party IME, which is a keylogger will get access to all the information the user entered. Then the keylogger selects the sensitive data it is interested in such as email and password and sends those data to its own server and stores the data in the database. 

\subsection{Third-Party IME}
With the popularity of touchscreen cell phones, unlike the traditional (physical) keyboard, people use the soft keyboard popped up on their touchscreen to input text. These soft keyboards are known as IME apps, and they convert user\rq s touch events to texts. Although there is always a built-in IME provided by the operating system (i.e. iOS or Android), many people tend to use a third-party IME app for better user experience, such as enhanced user interface, personal dictionary and extra emoji. However, at the same time they enjoy a third-party IME app, people are also exposed to a potential security threat. The third-party IME could be a malicious keylogger that stores your private sensitive information when you type in and send it over to a remote server or share it with another app. Two studies \cite{mohsen2016investigating}\cite{cho2015keyboard} of investigating the keylogging threat in Android show a high percentage of people use a third-party IME app on a daily basis and the high probability of being keylogged when using a third-party IME app. The high probability of keylogging comes from the large number of permissions a third-party IME has, such as INTERNET, READ\_CONTACTS, WRITE\_EXTERNAL\_STORAGE, RECORD\_AUDIO and READ\_PHONE\_STATE. Some of these permissions are used to provide enhanced features. For apps that require INTERNET permission, they may want to upgrade the app version with cloud server so the intention could be benign. However, INTERNET permission is potentially dangerous since it could be used to send whatever sensitive information to a remote server and user\rq s sensitive information would be leaked and it is entirely possible a malicious user may be able to collect such sensitive information and misbehave. Another permission that could be misused is write permission. An IME can write the sensitive information to a local storage and share the location of the file that contains the sensitive information with another app, then user\rq s sensitive information would be compromised. Therefore, it is important to have an effective way to prevent keylogging threat and make sure user\rq s sensitive information doesn't get leaked when using a third-party IME app. 

\subsection{Android OS}
Specifically, this paper focuses on Android platform since Android used to be the only mobile OS (operating system) that supports third-party IME. iOS didn\rq t support it until post iOS 8. In addition, Android currently has the largest market share in the world (70.85\% as in May 2016) \cite{marketshare}, so the research could have a wide impact. 

\subsection{Current Solutions}
There are a large number of papers studying about keylogging threats in computer. Most of the researches focused on software keyloggers in computer\cite{2011anti}\cite{aslam2004anti}\cite{fu2010detecting}\cite{shannon2006system} while some provided a solution to both software and hardware keyloggers in computer\cite{baig2007robust}. However, few of those approaches are applicable to mobile devices, since mobile device usually uses soft keyboard not physical keyboard, the underlying operating system is different and the computational power of mobile device is limited compared to computer. There are also a large number of papers studying about data leakage in mobile device, but they are not specific for keylogging threat, so those solutions are relatively complex and may not be effective for solving keylogging problem in mobile devices. The intersection of these two research areas is what this paper focuses on and this topic has not been fully studied. One latest study addressing keylogging threat in mobile device is I-BOX \cite{chen2015you}, but there are two limitations of the solution proposed in this study. One limitation is that there is a chance that sensitive data can be leaked in conjunction with another app. Another limitation is that it sometimes shuts down the internet when detecting sensitive data, which affects user experience. The motivation of my work is to address the limitation of I-BOX.

\section{KeyGuard}

\subsection{Threat Models}
We define threat models in three situations: 1) after keylogger IME obtains user\rq s keystroke event, keylogger IME directly sends keystroke event to a remote server; 2) keylogger IME logs user\rq s sensitive information and may misuse that information at a later stage; and 3) instead of sending keystroke event itself, keylogger IME colludes with another app. Keylogger IME stores the keystroke event to a local file and informs the colluded app to access that file, so the colluded app can read the file and send the data out or continue to collude with another app to pass the data down. 

This study mainly focuses on protecting sensitive data leakage through a third-party IME, i.e. security threats such as that after the user has entered his/her username and password in the app, the app leaks the sensitive information is out of the scope of this paper. Also, this paper attempts to identify and prevent the keylogging issue by monitoring keystroke events that are text typed through third-party IME soft keyboard, it does not tackle the security concerns that may rise from side channels such as voice input (voice-to-text conversion) or image input (sending pictures that is uploaded through third-party IME).

\subsection{Design Consideration}
There are four issues that are important to address in designing KeyGuard.
\begin{enumerate}
\item	The system should prevent the keylogging threat that is caused by storing user entered sensitive data in local storage, so that no other colluding app can obtain user\rq s sensitive data from accessing the local storage.
\item	The system should prevent the keylogging threat that is caused by sending user entered sensitive data to remote server through Internet.
\item	The system should maintain the functions of third-party IME when entering non-sensitive data, otherwise user cannot enjoy better user experience, which goes against the purpose of installing and using a third-party IME in the first place.
\item	The solution should not cause any noticeable effect to user experience, such as lag. 
\end{enumerate}

\subsection{Approach Overview}
The approach proposed is to intercept all keystroke events and only encrypt them if certain criteria are met before those events are handled by the IME, so the information stored by IME will be fundamentally different from what user has actually entered. Thus, this approach satisfies the first two design considerations in section 2.2, since the third-party IME can not touch the real data, it\rq s no longer a threat if third-party IME wants to store or send the data. 
The proposed solution works as follows:
\begin{enumerate}
\item It keeps monitoring all keystroke events sent to third-party IME.
\item If the keystroke event meets any of the following three conditions, perform the encryption and decryption operations mentioned below:
\begin{enumerate}
\item the input is coming from a password context (TYP\\E\_TEXT\_VARIATION\_PASSWORD)
\item the input matches predefined rules. For example, if the rule is defined as any string that starts with "abc" and has a total length of 8. After the user enters string "abc", the following 5 characters will be encrypted and decrypted so that third party IME doesn\rq t have access to them
\item the user performs a gesture and marks an input field as sensitive
\end{enumerate}
\item Every keystroke event that meets the sensitive condition above will be encrypted before it is handled by third-party IME.
\item The original text user entered will be decrypted when displaying on the underlying app.
\end{enumerate}

\subsection{User Model}
KeyGuard runs in the background but provides an user interface to allow users define sensitive information rules (e.g. any string that starts with "abc" and has a length of 8 characters). User uses third-party IME as usual, since this approach does not request user to use an additional keyboard or interface to enter their sensitive information, it saves the users from the hassle of switching back and forth between different keyboards etc. This approach does not need to limit the permission of third-party either, so user can enjoy the functions provided by third-party IME for all non-sensitive information, which satisfies the third design consideration in section 2.2.

\section{KeyGuard Design}
\subsection{Where to Intercept Keystroke Event/Restore Real Keystroke Event}
To decide where to intercept keystroke event, we first have to understand how user\rq s touch event on the soft keyboard translates to keystroke event and how that keystroke event is obtained by third-party IME. On a high level, the coordinates (x,  y) of the place where user touched are processed and a keyCode is determined through a mapping between screen locations and keyCodes. The keyCode is then passed on to key event listener for further handling. If we intercept the key event too early, we may not be able to know which key was actually entered, which in turn could affect the functionality of the third-party IME. If we intercept the key event too late, then the third-party IME could have already leaked the sensitive information. In order to implement the prototype, we assumed that the keylogging activity happened between \lstinline{KeyboardView.OnKeyboardActionListener.on Key(int primaryCode, int[] keyCodes)}, which is triggered whenever user enters a key, and \lstinline{InputMethodService.onUpdate Selection(int oldSelStart, int oldSelEnd, int newSelStart, int}\linebreak \lstinline{newSelEnd, int candidatesStart, int candidatesEnd)}, which is called when the application has reported a new selection region of the text. A study by L. Cai and H. Chen \cite{cai2011touchlogger} discussed the possibilities to infer actual keys pressed by users by collecting and analyzing the positions a user has touched on the soft keyboard. In theory any function that may leak sensitive information can also be hooked to mitigate the security threats. For the proof-of-concept prototype, we decided to go with the assumption made above.
\subsection{Only Encrypt Sensitive Data}
The purpose of keylogging is to obtain user\rq s sensitive information, so most of the research only take care of sensitive information. Similarly, in our study we would only need to encrypt sensitive data rather than encrypt everything. Our approach also followed this: we tried to encrypt information that appears in the sensitive input fields either through contexts such as password and email address, or those user has specified as containing sensitive information through gesture, and also encrypt strings matching user predefined sensitive information rules. By only encrypting sensitive information, we allow 3rd party IMEs to maintain their functionalities as many as possible so that end users\rq experience with their favorite IMEs will not degrade after using KeyGuard in mobile devices. 

\begin{table*}[h!]
  \centering
  \caption{Comparison of Systems}
  \label{tab:table1}
  \begin{tabular}{ccccc}
    \toprule
    Systems & Sensitive Data Protected & Add\rq l Windows & Functions Affected & Sensitive Info Exposed \\
    \midrule
    KeyGuard & all kinds & No & Virtually no noticeable impact  & Partial\\
    I-Box & all kinds & No & Internet temporarily disconnected & Partial\\
    ScreenPass & password only & Yes & Requires using a separate UI &Yes\\    
    Cashtag & all kinds & No & Requires entering aliases & Yes\\
    TinMan & all kinds & Yes & Requires using a separate UI & Yes\\
    \bottomrule
  \end{tabular}
\end{table*}

\subsection{Maintain Third-party IME Functionality}
Although there are a few solutions available to mitigate security concerns on leaking sensitive information in mobile devices, the biggest advantage of KeyGuard is that it allows the 3rd-party IME to maintain its functionalities as much as possible. We have compared five systems (including KeyGuard) in Table 1 and detailed description of the other four systems are available in Section 6. In terms of the amount of sensitive data protected, KeyGuard is on par with other solutions (except ScreenPass which is designed primarily for passwords) since it can protect all kinds of sensitive information as long as they are coming from a sensitive context, specified by user as containing sensitive information through gesture, and/or pre-defined in the rules. KeyGuard doesn\rq t require a new window to pop up just for the sake of entering sensitive information like ScreenPass and TinMan. When it comes to the number of 3-rd IME functions affected, unlike I-Box which requires Internet connection to be temporarily disconnected, or Cashtag which requires entering special alias as placeholders for sensitive information, KeyGuard introduces nearly no noticeable impact to user\rq s experience since it essentially performs no action for non-sensitive information. In the case that user specifies a field as containing sensitive information through gesture, it is probably fine to restrict some of the 3rd-party IME functionalities since it may not even be a desired experience to allow IME to perform word auto-suggest on sensitive information. Like I-Box, KeyGuard will expose partial sensitive information (rules that define sensitive information) to the 3rd-party IME, which is still more desirable than exposing all of sensitive information to the 3rd-party IME as in the cases of ScreenPass, Cashtag and TinMan.

\section{Implementation}
To verify our solution, we implemented a prototype based on the assumption mentioned in section 3. The hardware used to conduct this experiment is Sony Xperia Z1 tablet running on Android 4.2.2 Jelly Bean OS. The implementation is based on Xposed framework, which a number of researches in mobile securities have also adopted\cite{qiu2015apptrace}\cite{salman2014mobile}\cite{zhao2015privacy}. We used Google sample IME code\cite{googlesampleime} and modified it into a simple keylogger to serve as malicious third-party IME. 

\subsection{Xposed Framework}
Xposed\cite{xposedframework} is a framework for modules that can change the behavior of the system and apps without touching any APKs. The way apps work in Android is that, Android forks off a new virtual machine whenever an app is started. The Xposed framework allows overriding library routines to be inserted into the Java classpath through forking Zygote during app launching, prior to the execution of the new virtual machines. Thus, the overall system behavior is altered without modifying either the apps or the underlying firmware. Individual class methods can be hooked, allowing injected code to be executed prior to, following the completion of, or in place of the base-method calls. Private or protected member fields and functions can also be accessed and modified, and additional fields or functions can be added to the base class or object granularity.

\begin{figure}[ht]
  \graphicspath{{fig/}}
  \begin{center}
  \includegraphics[scale=0.3]{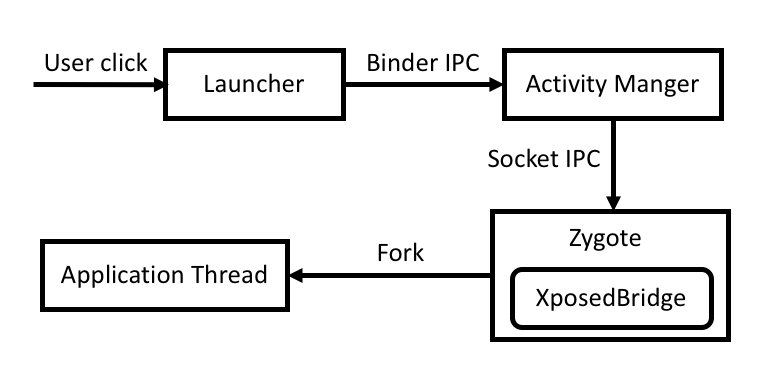} 
  \caption{Launch process of KeyGuard.}
  \label{fig:launch}
  \end{center}
  \end{figure}

\subsection{Google Sample IME}
In order to simulate the keylogging behavior of a third-party IME, we need to have the source code of a third-party IME. Due to the difficulty to find an open source commercial third-party IME, we used Google sample IME and modified the code to turn it into a keylogger. 

\begin{figure}
\centering
\includegraphics{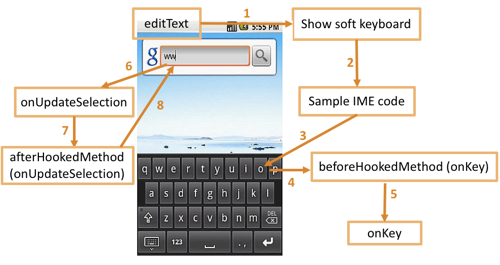}
\caption{The work flow of the KeyGuard prototype.}
\label{fig:workflow}
\end{figure}

\begin{figure}
\centering
\includegraphics{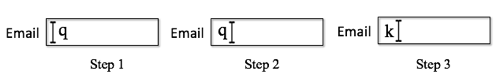}
\caption{Letter transformation when displayed in the underlying application.}
\label{fig:transform}
\end{figure}

\subsection{Functions Hooked}
The main functions that we hooked in our KeyGuard prototype are onKey() and onUpdateSelection() in the SoftKeyboard class. The work flow of KeyGuard prototype is shown in Figure \ref{fig:workflow}:

\begin{enumerate}[leftmargin=*]
\item User clicks on the textbox which invokes the input method service.
\item Google Sample IME (keylogger) serving as the current IME will be invoked.
\item Soft keyboard pops up and user types in a letter/number.
\item Code inside beforeHookedMethod() will be executed prior to onKey() to check if the keystroke entered meets the sensitive data condition. Only sensitive keystroke will be encrypted. RC4 has been adopted to leverage the speed of a stream cipher, but in theory any cipher could be used.
\item Keylogger executes onKey() method which gets the encrypted letter/number.
\item Encrypted letter/number is displayed on the textbox and cursor moves forward invoking the onUpdateSelection() method.
\item After onUpdateSelection() is executed, afterHookedMethod() method will be called.
\item Code inside afterHookedMethod() will be executed to decrypt the displayed letter/number to original one.
\end{enumerate}

Using Figure \ref{fig:transform} as an example, when user touches the editText control for entering password (this constitutes a sensitive context) on the screen, it invokes input method service and a soft keyboard pops up allowing user to enter password, when user touches a particular letter (k in this case) on the soft keyboard, but before the letter is received by the onKey() function in the third-party IME, the keyCode (letter) is encrypted (in beforeHookedMethod()) to be q before passing to the third-party IME, so q gets displayed on the screen for a short moment. When the cursor moves forward, onUpdateSelection() gets executed, after which the letter is decrypted (in afterHookedMethod()) and the original character is restored in the editText control in the underlying app.

\section{Evaluation}
\begin{figure}[ht]
  \graphicspath{{fig/}}
  \begin{center}
  \includegraphics[scale=0.45]{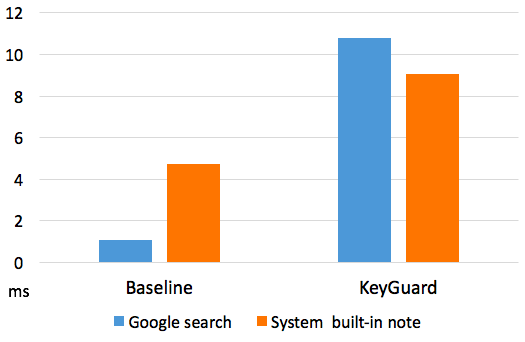} 
  \caption{Performance test.}
  \label{fig:performance}
  \end{center}
  \end{figure}

We conducted a performance test of our solution, specially around onKey() and  onUpdateSelection() functions. We did two rounds of performance tests, one in Google Chrome browser trying to enter search keyword; the other in system built-in note app. Over 60 key events have been performed in each test to generate a reasonable number of data points from which we calculated an average latency for onKey() and  onUpdateSelection() functions.The latency of each key event listener function increased after hooking the functions mentioned above. We are checking for sensitive data each time user typing in third-party IME and executing additional code before onKey() and after onUpdateSelection() for protecting sensitive data, so the increase in latency is understandable and expected. Though it looks like to be a significant increase in Figure \ref{fig:performance}, from a user experience perspective, the difference is negligible and the lag is hardly noticeable by users. The majority of users would not be able to tell a 10 ms increase in latency, so we are confident our solution can actually be deployed to end-users without causing performance degradation and user experience sacrifice on their mobile devices.

\section{Related Work}
TaintDroid \cite{enck2014taintdroid} uses a dynamic taint analysis to track the flow of sensitive information in order to determine whether sensitive information could be leaked through a third-party application to the network. But TaintDroid mainly focuses on detection, there is no further action to prevent privacy leakage. Besides, TaintDroid was designed to track sensitive information flow on an app-level. For malware written in pure C language or native code languages in android, it may not be effective in detecting/preventing attacks. But current IME apps tend to use excessive native code in their core logic, so TaintDroid based methods \cite{hornyack2011these} are hard to apply on this problem. 

I-BOX \cite{chen2015you}, an app-independent oblivious sandbox that significantly reduces chance of sensitive information leakage by confining untrusted IME apps to a set of predefined security policies. Specifically, I-BOX first maintains a checkpoint of an IME app\rq s state before any keystroke was received from the user. It keeps monitoring user\rq s keystroke events, and rolls the third-party IME back to its previous safe/clean state (checkpoint) if the user input matches predefined rules. One limitation of I-BOX is that an IME app could collude with another malware to leak information. An IME app could first save the user input in a local file, and inform a colluding malware to read the file when the transaction has not been rolled back and then divulge the input.

ScreenPass \cite{liu2013screenpass} employs a trusted software keyboard for inputting password. It will pop up a trusted password interface when user needs to enter password. Instead of typing password, user clicks the password tag. The ScreenPass system also adopts taint tracking to make sure that a password is exposed outside of a specific safe domain. Limitation of this method: 1) only works with password 2) all the passwords are stored in ScreenPass. If it is hacked, the attacker could get all the important information at once. Similar to the functionality of ScreenPass, TinMan \cite{xia2015tinman} also hides the plain text of password by making user select a password instead of typing password through keyboard. TinMan places confidential records on a trusted node and puts corresponding placeholders on the mobile device. But offloading from distributed shared memory causes big overhead. 

Cashtags\cite{mitchell2015cashtags} allows users to safely access pieces of sensitive information in public by intercepting and replacing sensitive data elements with non-sensitive data elements before they are displayed on the screen. Similar to ScreenPass, user has to provide all their sensitive information to the software in order to get their data replaced with a item begins with a cashtag alias, such as \$password. Whenever the sensitive term would be displayed on the screen, the system displays the predefined alias instead.This approach shares a similar idea of KeyGuard that is to intercept sensitive data, but this approach is not for keylogging, so the sensitive data it attempts to intercept are not keystroke events. Cashtags simply asks user to provide their sensitive data beforehand.

Cho et al \cite{cho2015keyboard} demonstrate the potential security risks by implementing a proof-of-concept keylogger that were able to effectively log users\rq sensitive keystrokes with 81 popular websites (out of 100 tested websites). They also analyzed the security behaviors of 139 keyboard applications that were available on Google Play. Their study results show that majority of existing keyboard applications (84 out of 139) could be potentially misused as malicious keyloggers.

\section{Discussion and Future Work}
One limitation of KeyGuard is that to use KeyGuard we need to intercept keystroke event and that involves modify system functions, so there is a must to require user's device to be rooted mainly to install the underlying Xposed framework. The root requirement could impact the vendor maintenance regime. One possible solution to this is to implement KeyGuard at the vendor level as an OEM (Original Equipment Manufacturer) feature.

Another limitation of KeyGuard is that the current prototype is based on the Google sample IME code which is open source. Xposed framework needs to know exactly which class and which method to hook for it to function properly, so it might be a challenge to extend the solution to other third-party IMEs since we may be able to view their source codes easily. Conceptually, the solution proposed in this paper should work with most of the third-party IME, however, we still need to further understand how to implement the idea of KeyGuard in different types of third-party IME.

Due to the time constraints, 2.(a) and 2.(b) mentioned in Section 3.3 have been implemented in the prototype. Allowing user to specify a field as containing sensitive information as mentioned in 2.(c) so that KeyGuard can kick in will be one of our future work as well.

There is another experiment we can carry out to see if it helps with obscuring sensitive information that is caught by the pattern detector. Instead of processing one character at a time, we could process characters every N milliseconds and we could see if that negatively affects user experience. This way if the user is typing something sensitive we could delay passing it to the IME just a little bit in hopes of getting more data to see if it should be encrypted so that the IME has less of a chance of seeing only partially encrypted data.

In addition, we will set up a usability study to determine how users perceive the ease of use of the system. The user study will help us find out what percentage of functionality in 3rd party IMEs doesn\rq t work and affects the overall user experience, using which we can further improve KeyGuard.

\section{Conclusions}
In order to solve the keylogging problem in mobile device, we proposed a novel solution to fundamentally solve the problem. Our approach resolves the concern that an IME can log down the sensitive information and send it to remote server or share it with another app by making IME store encrypted information so that the original information entered by user will not be visible to the third-party IME and thus it will be less likely to be leaked. The prototype demonstrates the feasibility and effectiveness of our approach. The evaluation of the prototype shows our solution will not introduce any user noticeable lag in terms of performance. These results suggest that our prototype succeeded in preventing keylogging threat and could be extended to third-party Android IME.

\bibliographystyle{abbrv}
\balance
\bibliography{KeyGuard} 
\end{document}